\documentclass[fleqn]{2023SCGE}
\usepackage{amsmath}
\usepackage{amssymb}
\usepackage{graphicx}
\usepackage{xcolor,color}
\usepackage{url}
\usepackage{bm}
\usepackage{mathrsfs}
\usepackage[utf8]{inputenc}
\usepackage{hyperref}
\usepackage{enumerate}
\usepackage{amsthm}
\usepackage{bbm}
\usepackage[normalem]{ulem}
\usepackage{upgreek}
\usepackage{tensor}
\usepackage{siunitx}
\usepackage{orcidlink}
\usepackage[caption=false]{subfig}
\usepackage{float}
\usepackage{colortbl}

\definecolor{colour1}{HTML}{0000ff} 

\usepackage{color,xcolor}
\usepackage{ulem}
\usepackage{float}
\usepackage{subfig}
\usepackage{cleveref}
\usepackage{multirow}
\usepackage{booktabs}
\usepackage{booktabs}
\usepackage{threeparttable}

\usepackage[export]{adjustbox}
\usepackage{times}
\usepackage{commath}

\crefname{equation}{Eq.}{Eq.}
\Crefname{equation}{Eqs.}{Eqs.}

\crefname{section}{Section}{Section}
\Crefname{section}{Sections}{Sections}

\crefname{appendix}{Appendix}{Appendix}

\crefname{table}{Table}{Table}
\Crefname{table}{Tables}{Tables}
\crefname{figure}{Figure}{Figure}
\Crefname{figure}{Figures}{Figures}

\begin{document}



\ArticleType{Article}
\Year{2024}
\Month{November }
\Vol{67}
\No{11}
\DOI{10.1007/s11433-024-2431-0}
\ArtNo{110411}
\title{Gravitational wave fluxes on generic orbits in near-extreme Kerr spacetime: higher spin and large eccentricity}{}

\author[1]{Changkai Chen\,\orcidlink{0000-0002-4023-0682}}{}%
\author[1]{{Jiliang} {Jing}\,\orcidlink{0000-0002-2803-7900}}{jljing@hunnu.edu.cn}

\address[1]{Department of Physics, Key Laboratory of Low Dimensional Quantum Structures and Quantum Control of Ministry of Education, \\Synergetic Innovation Center for Quantum Effects and Applications, Hunan Normal University, Changsha, 410081, Hunan, China}


\abstract{
To accurately obtain the waveform template of gravitational waves, substantial computational resources and extremely high precision are often required.
In a previous study [\textcolor{colour1}{JCAP {\bf 11} (2023) 070}], we employed the confluent Heun function to obtain an exact solution to the Teukolsky equation. This approach allowed us to efficiently and accurately calculate the gravitational wave flux for a particle in circular orbits around a Schwarzschild black hole.
Building on this method, we now extend its application to calculate the asymptotic gravitational wave fluxes from a particle in generic orbits around a near-extreme Kerr black hole. Our extended method proves particularly effective in handling computational challenges associated with large eccentricities ($e=0.9$), higher spins ($a=0.999$), higher harmonic modes, and strong-field regions. The results we obtained significantly outperform those derived from the numerical integration method based on the Mano--Suzuki--Takasugi method.

}



\maketitle



\begin{multicols}{2}
\section{Introduction}
The two-body problem is fundamental in general relativity, particularly in the field of gravitational wave (GW) astronomy. Binary black hole (BBH) inspirals, which generate GWs, have attracted significant attention and are expected to be directly detected by ongoing GW observatories worldwide, such as LIGO, Virgo, and KAGRA \cite{LIGOScientific:2016wkq,LIGOScientific:2016dsl,LIGOScientific:2018mvr,LIGOScientific:2020ibl,KAGRA:2021duu}.
Future space-based GW detectors, such as the Laser Interferometer Space Antenna (LISA) \cite{LISA:2017pwj,lisaconsortiumwaveformworkinggroup2023waveform}, TianQin \cite{TianQin:2015yph,TianQin:2020hid}, and Taiji \cite{Ruan:2018tsw}, are designed to detect GW signals from sources radiating in the millihertz bandwidth, notably extreme mass ratio inspirals (EMRIs).
EMRI events enable the precise mapping of black hole (BH) spacetimes, 
 \Authorfootnote

\noindent 
facilitating precise measurements of BH masses and spins.
 This offers an opportunity to test the hypothesis that astrophysical BHs conform to Kerr spacetime \cite{Babak:2017tow}.
Accurate predictions of GW waveforms are essential for detecting these signals in observational data.

A primary approach to achieving accurate waveform predictions is through calculations of gravitational self-force (GSF) within black hole perturbation (BHP) theory.
In this framework, the BBH system is modeled as a point mass orbiting a black hole, with its dynamics described by an equation of motion that includes the interaction with its self-field, known as the GSF.
The formalism for GSF calculations was first established by Mino, Sasaki, and Tanaka in the mid-1990s \cite{Mino:1996nk}, along with Quinn and Wald \cite{Quinn:1996am}. Over the past two decades, this formalism has been significantly refined to improve both its mathematical rigor and conceptual clarity.
For comprehensive reviews and further research on the GSF, please refer to Refs. \cite{Barack:2009ux, Poisson:2011nh, Wardell:2015kea, Pound:2015tma,Pound:2021qin}.

A commonly employed method for computing the GSF involves the Weyl scalars $\psi_0$ and $\psi_4$. which satisfy the separable Teukolsky equation \cite{Teukolsky:1972my,Teukolsky:1973ha,Press:1973zz,Teukolsky:1974yv}. These scalars encapsulate most of the gauge invariant information regarding the complete metric perturbation \cite{Wald:1973wwa}.
In the 1970s, Chrzanowski, Cohen, and Kegeles (CCK) \cite{Cohen:1974cm,Chrzanowski:1975wv,Kegeles:1979an} developed a technique to reconstruct vacuum metric perturbations in a radiation gauge from vacuum solutions of $\psi_0$ and $\psi_4$. However, as Ori later pointed out \cite{Ori:2002uv}, applying this method to the field generated by a point particle source results in highly singular metric perturbations.
These singularities occur not only at the particle location but also manifest as string-like gauge singularities extending from the particle to the event horizon or infinity.
It remains uncertain whether the established GSF form can be extended to account for the singularities associated with point particle sources. Pound et al. \cite{Pound:2013faa} later discovered that the GSF can indeed be extracted from the metric perturbations in the radiation gauge.
 Friedman's group \cite{Keidl:2006wk,Keidl:2010pm,Shah:2010bi} pioneered the use of the radiation gauge method for GSF calculations and determined the Detweiler redshift invariance for circular equatorial orbits in Kerr spacetime \cite{Shah:2012gu}. Maarten extended this method to calculate the first-order GSF and redshifts for eccentric equatorial orbits \cite{vandeMeent:2015lxa,vandeMeent:2016hel} and generic bound orbits\cite{vandeMeent:2017bcc}, which feature both eccentricity and inclination.
While $\psi_0$ or $\psi_4$ carry most of the information regarding metric perturbations, the perturbations related to ``mass'' and ``angular momentum'' must be recovered through alternative methods \cite{Wald:1973wwa}. Merlin et al. \cite{Merlin:2016boc} reconstructed these components for fields sourced by a particle in equatorial orbits by imposing the continuity of certain gauge-invariant fields derived from the metric.
Their results demonstrated simplicity: Outside the particle orbit, the mass and angular momentum perturbations are solely determined by the orbital energy and angular momentum, while both perturbations vanish within the orbit.
Using the direct method of metric perturbation generated by the CCK procedure, it was demonstrated that this result applies to any compact source in the radial direction \cite{vandeMeent:2017oet}.

Significant progress has been achieved in calculating the GSF for various orbits, but it remains particularly challenging for generic orbits, especially in Kerr spacetime. Generic geodesic orbits are characterized by three fundamental frequencies ${\Omega _{r,\theta,\phi }}$. However, calculating these frequencies is complex, and an efficient and accurate method for GSF calculations of generic orbits has yet to be developed.
Owing to the regularization problem at the point mass limit, high-precision GSF calculations demand substantial computational resources.
Reducing the computational cost of GSF calculations is therefore crucial.
The two-timescale expansion method \cite{Hinderer:2008dm} provides a hint: assuming that a point mass does not encounter any transient resonances \cite{Flanagan:2010cd}, the orbital phase becomes the most significant information for predicting the GW waveform. This can be expressed as follows:
\begin{equation}
\Phi = \epsilon^{-1} \left[ \Phi^{(0)} + \epsilon\Phi^{(1)} + O(\epsilon^2) \right],
\end{equation}
where the mass ratio $\epsilon = {m_2}/{m_1} \le 1$.
The leading-order term $\Phi^{(0)}$ can be considered as the time-averaged dissipative effect of the first-order GSF.
The calculation of the secular effect can be simplified by defining the radiative field as half the retarded solution of the gravitational perturbation equation minus half the advanced solution of the adiabatic approximation method \cite{Mino:2003yg,Sago:2005fn}. This approach is effective because the radiative field is a homogeneous solution to the divergence induced by the point mass limit. By using this method, the leading-order term can be accurately calculated without incurring huge computational costs.
On the other hand, $\Phi^{(1)}$ corresponds to the residual part of the first-order GSF. This includes the oscillatory components of both the dissipative and conservative GSFs, as well as the time-averaged dissipative part of the second-order GSF.
Currently, no simplified approach exists for accurately calculating these post-adiabatic components. However, since $\Phi^{(1)}$ is considered a minor factor, the requirement for precision is not stringent. This suggests that the computational load can be reduced by employing suitable methods with acceptable error tolerance to compute each piece of the GSF. Based on this formalism, Osburn et al. proposed a hybrid scheme for calculating the GSF of eccentric orbits \cite{Osburn:2014hoa}.

For non-spinning particles in generic orbits around a Kerr BH, the time-averaged dissipative part of the first-order GSF is the dominant contribution to inspiral evolution in Kerr spacetime, making it suitable for calculating GW fluxes necessary for constructing GW waveforms.
Using the Teukolsky equation to calculate the dissipative first-order GSF has led to significant breakthroughs.
Hughes successfully solved the short-range potential form \footnote{Sasaki--Nakamura gave a transformation \cite{SASAKI198185}, which converts the Teukolsky function $R(r)$ governed by an equation with long-range potential into the Sasaki--Nakamura function $X(r)$ governed by an equation with short-range potential.} of the Teukolsky equation by numerical integration \cite{Hughes:1999bq}, providing the inspiral evolution and snapshot waveforms for spherical orbits (circular and nonequatorial) in Kerr spacetime \cite{Hughes:2001jr}. Utilizing the frequency-domain orbit function of Kerr BHs \cite{Drasco:2003ky}, his group extended the GSF calculations to generic orbits\cite{Drasco:2005kz}.
After years of research, his group also produced long inspiral waveforms of EMRIs in generic orbits \cite{Hughes:2021}.
Moreover, they calculated the EMRI waveforms \cite{skoup2023asymptotic} for a spinning body moving on generic orbits up the second-order spin by using the frequency-domain method proposed in Refs.~\cite{Drummond:2022a,Drummond:2022b}.
Fujita's group from Japan first used the Mano--Suzuki--Takasugi (MST) method to develop an efficient numerical method for calculating the dissipative GSF of generic orbits \cite{Fujita:2009us}. Building on the MST method and the analytical solution for bound time-like geodesic orbits in Kerr spacetime \cite{Fujita:2009bp}, they derived high-order post-Newtonian (PN) expansion an analytical formula \cite{Sago:2015rpa}. This analytical formula was then used to study the measurability of EMRI waves by LISA, focusing primarily on the adiabatic waveforms of eccentric and equatorial sources \cite{Fujita:2020}. Their numerical and analytical methods are widely used in constructing waveforms for various mass ratios \cite{Isoyama:2021jjd,Wardell:2021fyy}.
The computational performance of these teams is exceptional. Their methods can accurately calculate the dissipative GSF for generic orbits in Kerr spacetime, even with large eccentricities, large inclinations, high harmonic modes, and strong-field regions. By contrast, other EMRI waveforms or GSF calculations are limited to small eccentricities ($e\leq 0.3$) or Schwarzschild BHs \cite{vandeMeent:2017bcc,Zhang:2022rfr,Zhang:2023vok}.
{Moreover, Yang and Han proposed a new general formalism for constructing GW waveforms in dirty EMRIs \cite{Jiang:2024lwg}.}
In our previous study \cite{chen2023exact}, we proposed an exact approach for calculating GW fluxes from a particle in circular orbits around a Schwarzschild BH using the confluent Heun function. The computational accuracy and efficiency of this exact approach are comparable to those of the MST and numerical integration methods.

In this work, {\it we extend the previous work to obtain GW radiative fluxes from a particle in generic orbits around a Kerr BH.} Our goal is to compute and store the grid data of radiative fluxes to cover the extreme cases of orbital parameters: $a = 0.999, e = 0.9,$ large inclinations and high harmonic modes. This work will provide an accurate and efficient choice of energy flux data and GW amplitudes for waveform construction.

The remainder of the paper is structured as follows:
\cref{sec:geodesics} provide a brief review of the geodesic motion of a point particle in Kerr spacetime.
In \cref{sec:Adiabaticevolution}, we present the exact solution of the Teukolsky equation of the Kerr BH expressed in the form of the confluent Heun function, which is used to calculate the radiative energy fluxes for generic orbits.
\cref{sec:Comparisons} compares our results with those from PN expansion and MST methods to validate the high precision of our approach.
Finally, \cref{sec:Conclusion} provides concluding remarks, including plans and directions for future work.
In this paper, we use geometrized units: $c=G=1$.


\section{Kerr geodesics}
\label{sec:geodesics}
In this section, we provide an overview of bounded geodesics in Kerr spacetime. For detailed content, please refer to Refs.\cite{Schmidt:2002qk,Drasco:2003ky,Drasco:2005kz,Fujita:2009bp,Sago:2015rpa,vandeMeent:2019cam,Hughes:2021}. Here, we briefly review the geodesic dynamics of a point particle in Kerr spacetime. Some lengthy but important formulas are provided in \ref{app:geodesicconstants}.

Firstly, the Kerr metric in the Boyer--Lindquist coordinates, $(t, r, \theta, \varphi)$, is given by
\begin{align}\label{eq:Kerr}
 g_{\mu\nu}dx^\mu dx^\nu &=-\left(1-\frac{2Mr}{\Sigma}\right)dt^2-\frac{4Mar\sin^2\theta}{\Sigma}dtd\varphi+\frac{\Sigma}{\Delta}dr^2\nonumber \\
&+\Sigma d\theta^2+\left(r^2+a^2+\frac{2Ma^2r}{\Sigma}\sin^2\theta\right)\sin^2\theta d\varphi^2,
\end{align}
where $\Sigma=r^2+a^2\cos^2\theta$, $ \Delta = ( {r - {r_-}} )( {r - {r_+}} )$ and ${r_\pm} = M \pm \sqrt {{M^2}-{a^2}}$. $r_- $ is the inner horizon, and $r_+ $ is the outer (event) horizon.
Here, $M$ and $aM$ denote the mass and angular momentum of the Kerr BH, respectively.

\subsection{Formulations of orbital frequencies}
We use Mino time as the time parameter to describe these bound orbits \cite{Fujita:2009bp,Hughes:2021}. An interval of Mino time $d\lambda$ is related to an interval of proper time using the equation $d\tau$ by $d\lambda = d\tau/\Sigma$. The geodesic equations become:
\begin{align}
\left(\frac{dr}{d\lambda}\right)^2 &= [E(r^2 + a^2) - aL_z]^2 - \Delta[r^2 + (L_z - aE)^2 + Q]
  \nonumber \\
& \equiv R(r),\label{eq:rdot}\\
\left(\frac{d\theta}{d\lambda}\right)^2 &= Q - \cot^2\theta L_z^2 - a^2\cos^2\theta(1 - E^2)\equiv\Theta(\theta)\;,\label{eq:thdot}\\
\frac{d\phi}{d\lambda} &= \csc^2\theta L_z + \frac{2Mr a E}{\Delta} - \frac{a^2 L_z}{\Delta}
\equiv\Phi_r(r) + \Phi_\theta(\theta)\;,\label{eq:phdot}\\
\frac{dt}{d\lambda} &= E\left[\frac{(r^2 + a^2)^2}{\Delta} - a^2\sin^2\theta\right] - \frac{2Mra L_z}{\Delta}
  \nonumber \\
&\equiv T_r(r) + T_\theta(\theta)\;.\label{eq:tdot}
\end{align}
where $E$, $L_z$, and $Q$ represent the orbit energy (per unit $\mu$), axial angular momentum (per unit $\mu$), and Carter constant (per unit $\mu^2$), respectively. The expressions of these quantities are shown in \Cref{eq:Eformula,eq:Lzformula,eq:Qformula}.

\Cref{eq:rdot,eq:thdot} show that the bound orbit of the radial and the polar motion is periodic when parameterized using $\lambda$:
\begin{equation}
r(\lambda) = r(\lambda + n\Lambda_r)\;,\quad
\theta(\lambda) = \theta(\lambda + k\Lambda_\theta)\;,
\end{equation}
where $n$ and $k$ are integers. For fundamental periods $\Lambda_{r,\theta,\phi}$, we can define the associated frequencies: $\Upsilon_{r,\theta,\phi} = 2\pi/\Lambda_{r,\theta,\phi}$.
The orbital motions in $t$ and $\phi$ can be decomposed into the long-term cumulative part and the oscillation functions:
\begin{align}
t(\lambda) &= t_0 + {\Upsilon _t}\lambda + \Delta t_r[r(\lambda)] + \Delta t_\theta[\theta(\lambda)]\;,\label{eq:t_of_lambda}\\
\phi(\lambda) &= \phi_0 + \Upsilon_\phi\lambda + \Delta\phi_r[r(\lambda)] + \Delta\phi_\theta[\theta(\lambda)]\;.
\label{eq:phi_of_lambda}
\end{align}
where $t_0$ and $\phi_0$ describe initial conditions,
\begin{equation}\label{eq:GammaUpsilon}
  {\Upsilon _t} = \langle T_r(r)\rangle + \langle T_\theta(\theta) \rangle\;, \quad \Upsilon_\phi = \langle \Phi_r(r)\rangle + \langle\Phi_\theta(\theta) \rangle\;,
\end{equation}
\begin{align}
  \Delta t_r[r(\lambda)] &= T_r[r(\lambda)] - \langle T_r(r)\rangle \equiv \Delta t_r(\lambda)\;,\\
 \Delta t_\theta[\theta(\lambda)] &= T_\theta[\theta(\lambda)] - \langle T_\theta(\theta)\rangle \equiv \Delta t_\theta(\lambda)\;,\label{eq:Deltat}\\
\Delta\phi_r[r(\lambda)] &= \Phi_r[r(\lambda)] - \langle \Phi_r(r)\rangle \equiv \Delta\phi_r(\lambda)\;,  \\
\Delta\phi_\theta[\theta(\lambda)] &= \Phi_\theta[\theta(\lambda)] - \langle \Phi_\theta(\theta)\rangle \equiv \Delta\phi_\theta(\lambda)\;.\label{eq:Deltaphi}
\end{align}

The quantities ${\Upsilon _t}$ and $\Upsilon_\phi$ represent the frequencies of coordinate time $t$ and $\phi$ with respect to $\lambda$, respectively.
The averages $\langle {T_{r,\theta}}\rangle$ and $\langle {\Phi_{r,\theta}}\rangle$ in \Cref{eq:GammaUpsilon,eq:Deltat,eq:Deltaphi} are defined as follows:
\begin{align}
\langle {T_r}(r)\rangle  &= \frac{1}{{{\Lambda _r}}}\int_0^{{\Lambda _r}} {{T_r}} [r(\lambda )]{\kern 1pt} d\lambda \;,
\\
 \langle {T_\theta }(r)\rangle & = \frac{1}{{{\Lambda _\theta }}}\int_0^{{\Lambda _\theta }} {{T_\theta }} [\theta (\lambda )]{\kern 1pt} d\lambda \;,\\
\langle {\Phi _r}(r)\rangle  &= \frac{1}{{{\Lambda _r}}}\int_0^{{\Lambda _r}} {{\Phi _r}} [r(\lambda )]{\kern 1pt} d\lambda \;,
\\
\langle {\Phi _\theta }(r)\rangle & = \frac{1}{{{\Lambda _\theta }}}\int_0^{{\Lambda _\theta }} {{\Phi _\theta }} [\theta (\lambda )]{\kern 1pt} d\lambda \;.
\end{align}

The observer-time frequencies can be expressed as follows:
\begin{equation}\label{eq:MTFreq}
{\Omega _{r,\theta ,\phi }} = \frac{{{\Upsilon _{r,\theta ,\phi }}}}{{\Upsilon _t} }.
\end{equation}

\subsection{Orbit parameterization and initial conditions}
\label{sec:params}
The ranges for radial and polar motion in the orbit are defined as $\theta_{\rm min} \le \theta \le \theta_{\rm max}$ and $r_{\rm min} \le r \le r_{\rm max}$, respectively. Here, $\theta_{\rm max} = \pi - \theta_{\rm min}$ and $r_{\rm min/max} = {p}/({1 \pm e})$.
To avoid ambiguity, we use the inclination angle $I$ instead of $\theta_{\rm min}$. The relationship between them is given by $I = \pi/2 - \mbox{sgn}(L_z)\theta_{\rm min}$. This allows for a smooth transition from $0$ for equatorial prograde orbits to $\pi$ for equatorial retrograde orbits.
The cosine of the inclination angle, $x_I \equiv \cos I$, serves as an excellent parameter to describe inclination: $x_I$ varies smoothly from $1$ to $-1$ as orbits vary from prograde equatorial to retrograde equatorial, with $L_z$ having the same sign as $x_I$.
A generic geodesic orbit in Kerr spacetime can be characterized by three parameters, $(E, L_z, Q)$. However, for bound orbits, it is often more convenient to use another set of parameters $(p, e, x_I)$ \cite{Schmidt:2002qk,vandeMeent:2019cam}.

\Cref{eq:t_of_lambda,eq:phi_of_lambda} provide the initial conditions $t_0$ and $\phi_0$.
To set initial conditions for the $r$ and $\theta$ directions, we introduce anomaly angles $\chi_r$ and $\chi_\theta$ to reparameterize these coordinate motions.

\begin{eqnarray}
r &=& \frac{p}{1 + e\cos(\chi_r + \chi_{r0})}\;,
\label{eq:rdef}\\
\cos\theta &=& \sqrt{1 - x_I^2}\cos(\chi_\theta + \chi_{\theta0})\;.
\label{eq:thetadef}
\end{eqnarray}
We set $\chi_\theta = 0$, $\chi_r = 0$, $t = t_0$, and $\phi = \phi_0$ when $\lambda = 0$.
The phase $\chi_{r0}$ then determines $r(0)$, and $\chi_{\theta0}$ determines the corresponding value of $\theta$.
Specifically, when $\chi_{\theta0} = 0$, the orbit has $\theta = \theta_{\rm min}$ when $\lambda = 0$; when $\chi_{r0} = 0$, it has $r = r_{\rm min}$ when $\lambda = 0$.

We define {\it the fiducial geodesic} as the geodesic with $\chi_{\theta0} = \chi_{r0} = \phi_0 = 0 = t_0$.
Quantities along the fiducial geodesic are denoted with a check mark.
For example, $\check r(\lambda)$ and $\check\theta(\lambda)$ represent the radial and polar directions along the fiducial geodesic.

For non-fiducial geodesics, we introduce the definitions: when $r = r_{\rm min}$, $\lambda_{r0} =\min(\lambda)$; when $\theta = \theta_{\rm min}$, $\lambda_{\theta0}=\min(\lambda)$.
This means that
\begin{equation}
r(\lambda) = \check r(\lambda - \lambda_{r0})\;,
\quad
\theta(\lambda) = \check \theta(\lambda - \lambda_{\theta0})\;,
\end{equation}
{These definitions show that these motions are periodic: $r(\lambda_{r0} + n\Lambda_r) = r_{\rm min}$ for any integer $n$, and $\theta(\lambda_{\theta0} + k\Lambda_\theta) = \theta_{\rm min}$ for any integer $k$.}
There are relationships between $\lambda_{\theta0}$ and $\chi_{\theta0}$, and between $\lambda_{r0}$ and $\chi_{r0}$. A useful corollary is
\begin{eqnarray}
\Delta t_r(\lambda) &=& \Delta\check t_r(\lambda - \lambda_{r0}) - \Delta\check t_r(-\lambda_{r0})\;,
\nonumber\\
\Delta t_\theta(\lambda) &=& \Delta\check t_\theta(\lambda - \lambda_{\theta0}) - \Delta\check t_\theta(-\lambda_{\theta0})\;,
\end{eqnarray}
with analogous formulas for $\Delta\phi_r$ and $\Delta\phi_\theta$.

Many of the definitions in this section are derived from Hughes' latest review of bound Kerr geodesics \cite{Hughes:2021}.
The introduction of these quantities $(\chi_{r0},\chi_{\theta0},\phi_0)$ facilitates the calculation of inspiral evolution in the adiabatic approximation.

\section{GW Radiative Fluxes via the Teukolsky equation}
\label{sec:Adiabaticevolution}

The next key step in constructing adiabatic inspirals is calculating the dissipative part of the first-order GSF. This can provide the associated gravitational wave amplitude and radiative fluxes. In this section, we extend the previous method \cite{chen2023exact} to Kerr spacetime to calculate these quantities.

\subsection{Exact Solutions of Teukolsky equations}\label{sec:solution_TE}
In the Teukolsky formalism, the gravitational perturbation of a Kerr black hole is described using the null-tetrad component of the Weyl tensors, specifically $\psi_0$ and $\psi_4$. These scalars satisfy the master equation \cite{Teukolsky:1973ha}.
The Weyl scalar $\Psi_4$ is related to the GW amplitude at infinity as follows:
\begin{equation}
  \psi_4\rightarrow\frac{1}{2}(\ddot{h}_{+}-i\,\ddot{h}_{\times }),\,\,\,{\rm for}\,\,\,r\rightarrow\infty.
\end{equation}
The master equation for $\psi_4$can be separated into radial and angular parts if we expand $\psi_4$ in Fourier harmonic modes as follows:
\begin{equation}\label{eq:psi4}
  \rho^{-4} \psi_4=\displaystyle \sum_{\ell m}\int_{-\infty}^{\infty} d\omega
e^{-i\omega t+i m \varphi} \ _{-2}S_{\ell m}^{a\omega}( \vartheta)
R_{\ell m\omega}(r),
\end{equation}
where $\rho=(r-i a \cos \vartheta)^{-1}$, and the angular function $_{-2}S_{\ell m}^{a\omega}( \vartheta)$ is the spin-weighted spheroidal harmonic \footnote{Note the distinction between $(\theta,\phi)$ of the orbits and $(\vartheta,\varphi)$ of the measured field. } with spin $s=-2$.
The radial function $R_{\ell m\omega}(r)$ satisfies the radial Teukolsky equation:
\begin{equation}\label{eq:GFoTRE}
 \left[ {\Delta ^{ - s + 1}\frac{d}{{dr}}\Delta^{s + 1}\frac{d}{{dr}} + V(r)} \right]{R_{\ell m\omega }} = {\Delta }{T_{\ell m\omega }},
\end{equation}
and the potential term $V(r)$ is expressed as:
\begin{equation}
   V(r) =  {K^2} - isK\Delta' + \Delta\left( {2isK' - \hat{\lambda} } \right),
\end{equation}
where $K=(r^2+a^2)\omega-ma$ and $\hat{\lambda}$ denotes the eigenvalue of $_{-2}S_{\ell m}^{a\omega}( \vartheta)$. The computational methods of the angular function $_{-2}S_{\ell m}^{a\omega}( \vartheta)$ and the source function ${T_{\ell m\omega }}$ are discussed in Refs. \cite{Hughes:1999bq,Sasaki:2003xr,OSullivan:2014ywd}.

The most commonly used methods for solving the homogenous solutions $R^{\rm in,up}_{\ell m\omega}$ of the Teukolsky equation are the PN expansion \cite{Tagoshi:1994sm,Tagoshi:1996gh,Tanaka:1996lfd,Shibata:1994jx} of the Sasaki--Nakamura equation and the MST method \cite{Sasaki:2003xr,Mano:1996mf,Mano:1996vt}.
We presented an exact approach that surpasses the MST method in terms of accuracy for calculating gravitational radiation at the horizon and infinity.
The key pieces of our approach have been summarized in depth in a previous paper \cite{chen2023exact}. Here, we provide a very brief sketch largely to set the context for the following discussion.
The general solution of the homogenous form of \Cref{eq:GFoTRE} is derived by the linear combination of two linearly independent particular solutions.
\begin{align}\label{eq:D2GSol1}
{R_{\ell m\omega }} &= {C_1}S_0^\beta (x){\rm{HeunC}}(\alpha ,\beta ,\gamma ,\delta ,\eta ;x) \nonumber\\
&+ {C_2}S_0^{ - \beta }(x){\rm{HeunC}}(\alpha , - \beta ,\gamma ,\delta ,\eta ;x).
\end{align}
where $C_1$ and $C_2$ represent constants that should be determined based on different boundary conditions. Here, ${\rm{HeunC}}$ denotes the confluent Heun function \cite{Ronveaux:1995,slavyanov2000special,olver2010nist}, and $x$ is defined as a new coordinate that is obtained by applying a M\"{o}bius (isomorphic) transformation, which is a linear fractional transformation of the form
\begin{equation}\label{eq:mtran}
x =- \frac{{  r - r_+}}{{r_ + } - {r_ - }},
\end{equation}
and the unnormalized S-homotopic transformation,
\begin{equation}\label{eq:D2SHT}
  S_{\rm{0}}^{\beta}(x) = {\left( { - x} \right)^{ \frac{1}{2}(\beta-s) }}{\left( {1 - x} \right)^{\frac{1}{2}(\gamma-s) }}{{\rm{e}}^{\frac{1}{2}\alpha x}},
\end{equation}
and the parameters $\alpha$, $\beta$, $\gamma$, $\delta$, and $\eta$ are provided by
\begin{subequations}
 \begin{align}
\alpha  &=    - 2i\omega {r_{\rm{x}}},\\
\beta  &= - s + {r_{\rm{x}}}^{ - 1}\left( {2i\omega (r_ + ^2 + {a^2}) - 2iam} \right),\\
\gamma  &= s + {r_{\rm{x}}}^{ - 1}\left( {2i\omega (r_ - ^2 + {a^2}) - 2iam} \right),\\
\delta  &=  2\omega {r_{\rm{x}}}(\omega ({r_ - } + {r_ + }) + is),\\
\eta  &=  2is\omega {r_ + } - \frac{1}{2}{s^2} - s - \hat{\lambda} \nonumber \\
& - 2r_{\rm{x}}^{ - 2}(\omega {r_{\rm{m}}} - am)(\omega {\hat r_{\rm{o}}} - am).
\end{align}
\end{subequations}
where ${r_{\rm{x}}} = {r_ - } - {r_ + },\hat{r}_{\rm o}={a^2} + 2{r_-} {r_+} - r_{+}^2$, and $r_{\rm m} = r_+ ^2+{a^2}$.
A more detailed derivation of the general solution \eqref{eq:D2GSol1} of the homogeneous Teukolsky equation can be seen in our previous work \cite{chen2023exact}.
However, this derivation did not include the ingoing wave and outgoing wave solutions for the Kerr spacetime and their asymptotic amplitudes. Next, we will present the analytical expressions for these solutions.

The selection of the general solution \eqref{eq:D2GSol1} is advantageous in constructing solutions that satisfy the normalized boundary conditions of the pure ingoing wave $R^{\rm in}_{\ell\omega}(r)$ at the horizon and the pure outgoing wave $R^{\rm up}_{\ell\omega}(r)$ at infinity. Specifically:
\textcolor{red}{
\begin{align}
    & R_{\ell m\omega }^{{\rm{in}}}\to \left\{ {
    \begin{array}{*{20}{l}}
{{\Delta ^{ - s}}{{\rm{e}}^{ - iP{r^*}}},}&{r \to r_{+}},\\
{{{B_{\ell m\omega }^{{\rm{ref}}}}}{r^{-1-2s}}{{\rm{e}}^{i\omega {r^*}}} + {{B_{\ell m\omega }^{{\rm{inc}}}}}{r^{-1}}{{\rm{e}}^{ - i\omega {r^*}}},}&{r \to  + \infty ,}
\end{array}
    } \right.\label{eq:boundary1}\\
 & R_{\ell m\omega }^{{\rm{up}}} \to \left\{ {\begin{array}{*{20}{l}}
{C_{\ell m\omega }^{{\rm{up}}}{{\rm{e}}^{iP{r^*}}} + {C_{\ell m\omega }^{{\rm{ref}}}}{\Delta ^{-s}}{{\rm{e}}^{ - iP{r^*}}},}&{r \to {r_{+} },}\\
{{r^{-1-2s}}{{\rm{e}}^{i\omega {r^*}}},}&{r \to  + \infty ,}
\end{array}} \right.\label{eq:boundary2}
\end{align}
}
where ${P}=\omega-m\Omega_{\rm H} $ and $\Omega_{\rm H} = \frac{a}{2Mr_+}$. Here, $r^*$ is the tortoise coordinate defined by
\begin{equation}
 {r^*} = r + \frac{{2M{r_ + }}}{{{r_ + } - {r_ - }}}\ln \frac{{r - {r_ + }}}{{2M}} - \frac{{2M{r_ - }}}{{{r_ + } - {r_ - }}}\ln \frac{{r - {r_ - }}}{{2M}}.
\end{equation}

To construct the two solutions $R_{\ell m\omega }^{{\rm{in,up}}}$ that satisfy the boundary conditions, we need to determine the coefficients $C_1$ and $C_2$ in the general solution \eqref{eq:D2GSol1}. Prior to undertaking this task, it is essential to determine the asymptotic expressions of the confluent Heun function near the horizon and at infinity.

Expanding the confluent Heun function around the irregular singular points at event horizon $x=0$ and infinity $x=\infty$, the asymptotic behavior at infinity can be expressed as:
\begin{subequations}\label{eq:hc-hor-inf}
\begin{align}
& \mathop {\lim }\limits_{x \to 0} {\rm{ HeunC}}(\alpha ,\beta ,\gamma ,\delta ,\eta ;x) = 1,\\
&\mathop {\lim }\limits_{\left| x \right| \to \infty } {\rm{HeunC}}(\alpha ,\beta ,\gamma ,\delta ,\eta ;x) \to D_ \odot^\beta \;{x^{ - \frac{{\beta  + \gamma  + 2}}{2} - \frac{\delta }{\alpha }}} + D_ \otimes ^\beta {{\rm{e}}^{ - \alpha x}}{x^{ - \frac{{\beta  + \gamma  + 2}}{2} + \frac{\delta }{\alpha }}},
\end{align}
\end{subequations}
where ${{  D}_{\otimes}^\beta }$ and ${{  D}_{\odot }^\beta }$ are constants, and their analytical expressions are shown in \ref{app:AsymptoticFormula}.

Using the asymptotic properties \eqref{eq:hc-hor-inf} of the confluent Heun function and the boundary conditions \eqref{eq:boundary1} of the ingoing wave, we can construct the ingoing wave solution.
The ingoing wave solution at the horizon has a purely ingoing property, which implies $C_2^{\rm in}=0$. Thus, the ingoing wave solution $R_{\ell m\omega }^{{\rm{in}}}$ is given by:
\begin{equation}\label{eq:uHor}
R_{\ell m\omega }^{{\rm{in}}} = S_0^\beta (x) {\rm{HeunC}}(\alpha ,\beta ,\gamma ,\delta ,\eta ;x).
\end{equation}

By utilizing the asymptotic behavior \eqref{eq:boundary1} of the solution $ R_{\ell m\omega }^{{\rm{in }}}$ as $r\rightarrow \infty$, we derive analytic expressions for the asymptotic amplitudes $B_{\ell m\omega }^{\rm inc}$ and $B_{\ell m\omega }^{\rm ref}$,
\begin{subequations}
\begin{align}
 &B_{\ell m\omega }^{{\rm{inc}}} = {( - {r_{\rm{x}}})^{1 - 2s + i\frac{{ma}}{{{r_ + }}}}}{\left( {2M} \right)^{ - i\frac{{ma}}{{{r_ + }}}}}{{\rm{e}}^{i\frac{{ma}}{{2M}}}}{( - 1)^{ - \frac{{\beta  + \gamma  + 2}}{2} - \frac{\delta }{\alpha }}}\;D_ \odot ^\beta ,\\
&B_{\ell m\omega }^{{\rm{ref}}} =( - {r_{\rm{x}}}){\left( { - \frac{{2M}}{{{r_{\rm{x}}}}}} \right)^{2iM(P + \omega )}}{{\rm{e}}^{ - i(P + \omega ){r_ + }}}{( - 1)^{ - \frac{{\beta  + \gamma  + 2}}{2} + \frac{\delta }{\alpha }}}D_ \otimes ^\beta.
\end{align}
\end{subequations}

Similarly, by utilizing the asymptotic properties \eqref{eq:hc-hor-inf} of the HeunC function and the boundary conditions \eqref{eq:boundary2} for the outgoing wave, we can construct the outgoing wave solution, which can be given by
\begin{align}\label{eq:uOut1}
  R_{\ell m\omega }^{{\rm{up}}}=&\Big[ {\left( { - 1} \right)^{\beta+1} } {  \frac{{D_ \odot ^{ - \beta }}}{{D_ \odot ^\beta }}} S_0^\beta (x)  {\rm{HeunC}}(\alpha ,\beta ,\gamma ,\delta ,\eta ;x)
 \nonumber \\
& + S_0^{-\beta} (x) {\rm{HeunC}}(\alpha ,-\beta ,\gamma ,\delta ,\eta ;x)\Big].
\end{align}

By utilizing the asymptotic behavior \eqref{eq:boundary2} of the solution $R_{\ell m\omega }^{{\rm{up }}}$ as $r\rightarrow r_{\rm{H}}$, we derive analytic expressions for the asymptotic amplitudes $C_{\ell m\omega }^{{\rm{ref}}}$ and $C_{\ell m\omega }^{{\rm{up}}}$,
\begin{subequations}
  \begin{align}
&C_{\ell m\omega }^{{\rm{ref}}} ={( - {r_{\rm{x}}})^{ - 1 + 2iM(P + \omega )}} {(2M)^{ - 2iM(P + \omega )}}
 {{\rm{e}}^{i(P + \omega ){r_ + }}}\nonumber \\
&\times {( - 1)^{2 + \frac{{\beta  + \gamma }}{2} - \frac{\delta }{\alpha }}}\frac{{D_ \odot ^{ - \beta }}}{{D_ \odot ^\beta }}\tilde D,\\
&C_{\ell m\omega }^{{\rm{up}}} = {( - {r_{\rm{x}}})^{ - 2s - 1 + i\frac{{ma}}{{{r_ + }}}}}{(2M)^{ - i\frac{{ma}}{{{r_ + }}}}}{{\rm{e}}^{i\frac{{ma}}{{2M}}}}{( - 1)^{\frac{{ - \beta  + \gamma  + 2}}{2} - \frac{\delta }{\alpha }}}\tilde D.
\end{align}
\end{subequations}
with
\[\tilde D = {\left( {D_ \otimes ^{ - \beta } - \frac{{D_ \odot ^{ - \beta }}}{{D_ \odot ^\beta }}D_ \otimes ^\beta } \right)^{ - 1}}.\]

The parameters $(\alpha,\beta,\gamma,\delta,\eta)$ of the confluent Heun function in this paper use the notation from the \textit{Maple} software. To facilitate readers who may use other software for calculations, we provide the conversion relationships for the parameters of the confluent Heun function between various software packages in \ref{app:HeunCParameters}.

\subsection{Radiative Fluxes}
The separated radial function $R_{\ell m\omega }(r)$ demonstrates simple asymptotic behaviors that are only ingoing at the horizon and only outgoing at infinity.

\begin{equation}
  {R_{\ell m\omega }}(r) \to \left\{ \begin{array}{l}
Z_{lm\omega }^\infty {r^3}{e^{i\omega {r^*}}}\;,\qquad r \to \infty \;,\\
Z_{lm\omega }^{\rm{H}}\Delta {e^{ - iP{r^*}}}\;,\qquad r \to {r_ + }\;.
\end{array} \right.
\end{equation}

The amplitudes $Z^{\infty,{\rm H}}_{lm\omega}$ are obtained using the Green's function method, which involves integrating homogeneous solutions $R_{\ell m\omega }^{{\rm{in,up}}}$ and the source term ${T_{\ell m\omega }}$ from the separated radial Teukolsky equation.
For a detailed discussion on non-rotating BHs, please refer to Ref.\ \cite{chen2023exact}.
The integral formula of the amplitudes $Z^{\infty,{\rm H}}_{lm\omega}$ can be written in the following form:
\begin{equation}
Z^{\infty,{\rm H}}_{lm\omega} = \int_{-\infty}^\infty d\tau\,e^{i\omega [t(\tau) - t_0]} e^{-im\phi(\tau)}{\cal I}^{\infty,{\rm H}}_{lm\omega}[r(\tau), \theta(\tau)]\;.
\label{eq:Zlmw1}
\end{equation}
where the integration variable $\tau$ is proper time along the geodesic.

The function ${\cal I}^{\infty,{\rm H}}_{lm\omega}(r,\theta)$ is discussed in our previous work \cite{chen2023exact}. This function is schematically a Green's function used to obtain ingoing wave and outgoing wave solutions ${R_{\ell m\omega }^{{\rm{in,up}}}}$ of the Teukolsky equation, multiplied by the source term of this equation.
\begin{align}
 {\cal I}^{\infty,{\rm H}}_{lm\omega}&=  \frac{1}{{W_{\rm C}}}\left[ {\left( {{A_{nn0}} + {A_{\bar mn0}}+{A_{\bar m\bar m0}}}\right)R_{\ell m\omega }^{{\rm{in,up}}}} \right. \nonumber \\
 &- \left( {{A_{\bar mn1}} + {A_{\bar m\bar m1}}} \right){\left( {R_{\ell m\omega }^{{\rm{in,up}}}} \right)^\prime }{\left. + {A_{\bar m\bar m2}}{{\left( {R_{\ell m\omega }^{{\rm{in,up}}}} \right)}^{\prime \prime }} \right]},
\end{align}
where $'$ denotes $\partial/\partial r$, and ${{\rm{W}}_C}$ is the conserved Wronskian, that is
\begin{equation}
   {W_{\rm C}} = R_{\ell m\omega }^{{\rm{up}}}\frac{d}{{d{r^*}}}R_{\ell m\omega }^{{\rm{in}}} - R_{\ell m\omega }^{{\rm{in}}}\frac{d}{{d{r^*}}}R_{\ell m\omega }^{{\rm{up}}}= 2i\omega B_{\ell m\omega }^{{\rm{inc}}}.
\end{equation}
and $A_{nn0}$ and other terms are given in Refs. \cite{Sasaki:2003xr,Teukolsky:1973ha}.
From \Cref{eq:t_of_lambda,eq:phi_of_lambda}, we can change the integration variable from proper time $\tau$ to Mino time $\lambda$ for \Cref{eq:Zlmw1}.
\begin{equation}
Z^{\infty,{\rm H}}_{lm\omega} = e^{-i m\phi_0}\int_{-\infty}^\infty d\lambda\, e^{i(\omega{\Upsilon _t} - m\Upsilon_\phi)\lambda} J^{\infty,{\rm H}}_{lm\omega}[r(\lambda),\theta(\lambda)]\;,
\label{eq:Zlmw2}
\end{equation}
where we have introduced
\begin{align}
J^{\infty,{\rm H}}_{lm\omega}(r,\theta) &= (r^2 + a^2\cos^2\theta){\cal I}^{\infty,{\rm H}}_{lm\omega}(r,\theta)\nonumber \\
&\times e^{i\omega\left[\Delta t_r(r) + \Delta t_\theta(\theta)\right]} e^{-im\left[\Delta\phi_r(r) + \Delta\phi_\theta(\theta)\right]}\;.
\end{align}

Owing to the periodicity of $r$ and $\theta$ directions of orbital motions with respect to Mino time, the function $J^{\infty,{\rm H}}_{lm\omega}$ can be expressed as a double Fourier series \cite{Hughes:2021}:
\begin{equation}
J^{\infty,{\rm H}}_{lm\omega} = \sum_{k=-\infty}^\infty\sum_{n = -\infty}^\infty J^{\infty,{\rm H}}_{\ell mkn} e^{-i(k\Upsilon_\theta + n\Upsilon_r)\lambda}\;,
\label{eq:Jlmwexpand}
\end{equation}
where
\begin{align}
J^{\infty,{\rm H}}_{\ell mkn} &= \frac{1}{\Lambda_r\Lambda_\theta}\int_0^{\Lambda_r}d\lambda_r\,e^{in\Upsilon_r\lambda_r}\nonumber \\
&\times \int_0^{\Lambda_\theta}d\lambda_\theta\,e^{ik\Upsilon_\theta\lambda_\theta}\,J^{\infty,{\rm H}}_{lm\omega}[r(\lambda_r),\theta(\lambda_\theta)]\;.
\label{eq:Jlmkn_def}
\end{align}

From \Cref{eq:Zlmw2,eq:Jlmwexpand,eq:Jlmkn_def,eq:MTFreq}, we find that
\begin{equation}
Z^{\infty,{\rm H}}_{lm\omega} = \sum_{k = -\infty}^\infty\sum_{n = -\infty}^\infty Z^{\infty,{\rm H}}_{\ell mkn}\delta(\omega - \omega_{mkn})\;,
\label{eq:ZlmknZlmw}
\end{equation}
where
\begin{equation}
\omega_{mkn} = m\Omega_\phi + k\Omega_\theta + n\Omega_r
\end{equation}
and
\begin{equation}
Z^{\infty,{\rm H}}_{\ell mkn} = e^{-i m\phi_0}J^{\infty,{\rm H}}_{\ell mkn}/{\Upsilon _t}\;.
\label{eq:Zlmkn_geod}
\end{equation}
These coefficients demonstrate the symmetry
\begin{equation}
Z^{\infty,{\rm H}}_{\ell,-m,-k,-n} = (-1)^{(\ell+k)}\bar Z^{\infty,{\rm H}}_{\ell mkn}\;,
\label{eq:Zlmkn_sym}
\end{equation}
where the overbar denotes complex conjugation.

In the actual calculation of the amplitudes $Z^{\infty,{\rm H}}_{\ell mkn}$, we can take advantage of the symmetry \eqref{eq:Zlmkn_sym} properties of the amplitudes $Z^{\infty,{\rm H}}_{\ell mkn}$ to reduce computational effort. We first calculate the amplitudes $Z^{\infty,{\rm H}}_{\ell mkn}$ for all $\ell$, all $m$, all $k$, and $n \ge 0$, to calculate the half of the modal amplitudes, and then use \Cref{eq:Zlmkn_sym} to obtain the remaining amplitudes.
The efficient calculation of the amplitudes $Z^{\infty,{\rm H}}_{\ell mkn}$ enables us to quickly obtain the GW radiative fluxes and the inspiral waveform of adiabatic evolution.

The phase of $Z^{\infty,{\rm H}}_{\ell mkn}$ depends on the initial values of $(\lambda_{r0},\lambda_{\theta0})$, which in turn depend on the initial anomaly angles $(\chi_{r0},\chi_{\theta0})$. We use $\check Z^{\infty,{\rm H}}_{\ell mkn}$ to represent the amplitudes $Z^{\infty,{\rm H}}_{\ell mkn}$ of the fiducial geodesic. Consequently, the new expression of $Z^{\infty,{\rm H}}_{\ell m\omega}$ can be written as a phase factor multiplied by the fiducial geodesic value $Z^{\infty,{\rm H}}_{\ell m\omega}$,
\begin{equation}
Z^{\infty,{\rm H}}_{\ell mkn} = e^{i\xi_{mkn}}\check Z^{\infty,{\rm H}}_{\ell mkn}\;,
\end{equation}
where the correcting phase is
\begin{align}
\xi_{mkn} &= k\Upsilon_\theta\lambda_{\theta0} + n\Upsilon_r\lambda_{r0}
+ m\left[\Delta\check\phi_r(-\lambda_{r0}) + \Delta\check\phi_\theta(-\lambda_{\theta0}) - \phi_0\right]
\nonumber\\
&-  \omega_{mkn}\left[\Delta\check t_r(-\lambda_{r0}) + \Delta\check t_\theta(-\lambda_{\theta0})\right]\;.
\label{eq:ximkn}
\end{align}

The initial condition affects these amplitudes only through the phase factor $\xi_{mkn}$. This fact means that we only need to calculate and store the quantities for the fiducial geodesics. By using \Cref{eq:ximkn}, we can easily convert these results to any initial conditions. This greatly reduces the amount of computation required to cover physically important EMRI systems. Therefore, $\xi_{mkn}$ can be rewritten in the following form \cite{Hughes:2021}:
\begin{equation}
\xi_{mkn} = m\xi_{100} + k\xi_{010} + n\xi_{001}\;.
\end{equation}
For each orbit, one needs to compute only ($\xi_{100},\xi_{010},\xi_{001}$) to know the phases for all $(\ell,m,k,n)$.

{
Next, to obtain the radiative fluxes of energy and angular momentum, we just multiply the amplitudes of each mode by its weight coefficient and then sum it.
\begin{equation}
  \left\langle {\frac{{dE}}{{dt}}} \right\rangle  = \sum\limits_{\ell mkn} {\frac{1}{{4\pi \omega _{mkn}^2}}} \left( {|Z_{\ell mkn}^\infty {|^2} + {\alpha _{\ell m\omega }}|Z_{\ell mkn}^{\rm{H}}{|^2}} \right)\;\;,\label{eq:dEdt}
\end{equation}
\begin{equation}
   \left\langle {\frac{{d{L_z}}}{{dt}}} \right\rangle  = \sum\limits_{\ell mkn} {\frac{m}{{4\pi \omega _{mkn}^3}}} \left( {|Z_{\ell mkn}^\infty {|^2} + {\alpha _{\ell m\omega }}|Z_{\ell mkn}^{\rm{H}}{|^2}} \right)\;,\label{eq:dLzdt}
\end{equation}
where $\left<\cdots\right>$ represents the time average. And
\begin{align}
  \alpha_{\ell m\omega} &=
 \frac{256(2Mr_+)^5 P (P ^2+4\tilde\epsilon^2)(P ^2+16\tilde\epsilon^2)\omega^3}
{|{\bf{C}}|^2},\\
   |{\bf{C}}|^2&= (2\,\lambda+3)\,(96\,a^2\,\omega^2-48\,a\,\omega\,m)+144\,\omega^2\,(M^2-a^2) \nonumber\\
       +& \left[(\lambda+2)^2+4\,a\,\omega\,m-4\,a^2\,\omega^2\right]\left[\lambda^2+36\,a\,\omega\,m-36\,a^2\,\omega^2\right].\nonumber\\
\end{align}
with $\tilde\epsilon=\sqrt{M^2-a^2}/(4Mr_{+})$.
}

For simplicity, we define some abbreviations:
\begin{align}
&\{ \left\langle {dE/dt} \right\rangle ,\left\langle {d{L_z}/dt} \right\rangle \}  = \{ \dot E,{{\dot L}_z}\} ,\\
&\sum\limits_{\ell mkn} {}  \to \sum\limits_{\ell  = 2}^\infty  {\sum\limits_{m = 1}^\ell  {\sum\limits_{k =  - \infty }^\infty  {\sum\limits_{n =  - \infty }^\infty  {} } } }.
\end{align}

When the object is in a Kerr geodesic orbit, it can be characterized by the orbital integrals $E$ and $L_z$ (under initial conditions).
The results of $ \{ \dot E,{\dot L_z}\}$ have been known for a long time \cite{Teukolsky:1974yv};
It can be realized from \Cref{eq:dEdt,eq:dLzdt} that these quantities can be understood as two regular fields, which remain regular at the event horizon and infinity, respectively.
The former can be calculated by the gravitational radiation at infinity.
Calculating the gravitational radiation at the event horizon is a bit tricky; you have to calculate how tidal stresses shear generators of the horizon, increase its surface area (or entropy), and then apply the first law of BH dynamics to infer changes in the mass and spin of BHs.

\section{Comparisons with other methods}\label{sec:Comparisons}
In our previous papers \cite{chen2023exact}, we demonstrated that the energy flux calculation using the HeunC method significantly outperforms both high-order post-Newtonian expansions \cite{Fujita:2014eta} and the numerical MST method \cite{BHPToolkit,Fujita:2004rb,Fujita:2005kng} for particles in circular orbits around non-rotating BHs. In this section, we numerically simulate the motion of test particles around rotating BHs along generic orbits and compare the computational performance of various methods \cite{BHPToolkit}.
{
 For the calculation of energy fluxes using the BHPToolkit \cite{BHPToolkit}, a numerical integration method\footnote{
 As described in Refs. \cite{Gralla:2015rpa,BHPToolkit}, the Teukolsky equation is first transformed into Sasaki--Nakamura equation, and then the numerical integration method is used to solve this equation according to the specific boundary conditions of the Sasaki--Nakamura equation. However, this method cannot solve the asymptotic amplitudes $B_{\ell m\omega }^{{\rm{inc,ref}}}$ and $C_{\ell m\omega }^{{\rm{inc,up}}}$ of the homogeneous Teukolsky equation.} is used to calculate the homogeneous Teukolsky equation, and the full MST method is used to calculate the asymptotic amplitudes $B_{\ell m\omega }^{{\rm{inc,ref}}}$ and $C_{\ell m\omega }^{{\rm{inc,up}}}$. By combining the MST method with numerical integration, we can calculate \Cref{eq:Zlmw2} more efficiently.
 }
We refer to this hybrid approach as the MST-NI method. In this paper, we present exact solutions using the MST-NI method, with floating-point numbers ${\bf N} =300$. Here, the floating-point numbers ${\bf N}$ correspond to ``MachinePrecision" in Mathematica, which affects computational efficiency. Higher values of $\bf N$, th require more computation time.
{In our tables and figures, the relative errors of the radiative flux $\dot{E}$ are the numerical round-off errors, and their definition is
\begin{equation}
  \Delta \dot E = \left| {\frac{{{{\dot E}_{{\rm{Exact}}}} - {{\dot E}_0}}}{{{{\dot E}_{{\rm{Exact}}}}}}} \right|
\end{equation}
where ${{\dot E}_{{\rm{Exact}}}}$ represents the energy flux of the exact solution, and ${{\dot E}_0}$ denotes the energy flux of a certain method.
}

\begin{figure*}[htbp]
	\centering
\includegraphics[width=7in]{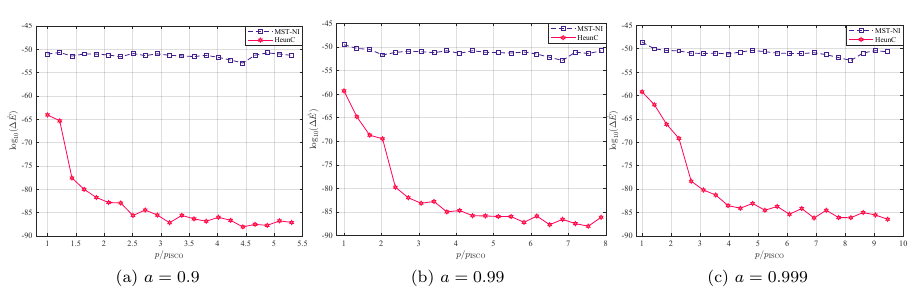}
\caption{Logarithmic relative errors of radiative fluxes ($\ell_{\rm max}=6$) of circular orbits at $p=[p_{\rm ISCO},p_{\rm ISCO}+10M]$ with large spins and ${\bf{N}}=100$ (Figures (a), (b) and (c) show spins equal to 0.9, 0.99 and 0.999 respectively). }
\label{fig:FluxCircular}
\end{figure*}
\begin{figure*}[htbp]
	\centering
\includegraphics[width=7in]{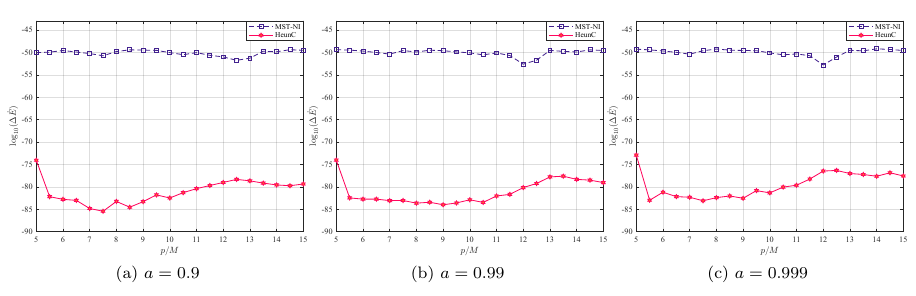}
\caption{Logarithmic relative errors of radiative fluxes ($\ell_{\rm max}=6$ and $1 \le  k \le 5$) of spherical orbits ($x_{I}=0.9$) at $p=[5M,15M]$ with large spins and ${\bf{N}}=100$ (Figures (a), (b) and (c) show spins equal to 0.9, 0.99 and 0.999 respectively). }
\label{fig:FluxSpherical}
\end{figure*}

\begin{figure*}[htbp]
	\centering
\includegraphics[width=7in]{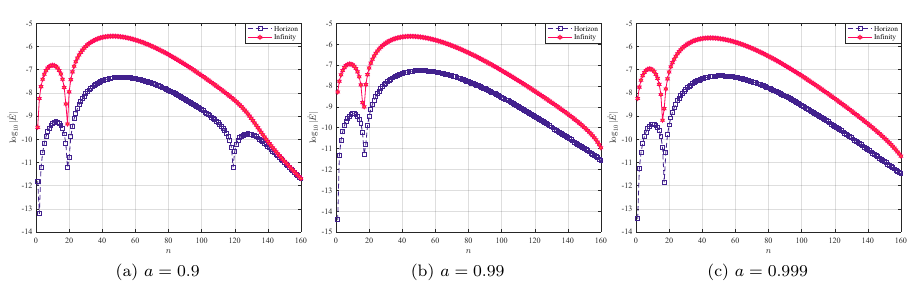}
\caption{Modal radiative fluxes ($\ell= m = 2$) through the horizon and at infinity for the equatorial eccentric orbit ($e = 0.9$) at $p = 6M$ with large spins  (To obtain the data quickly, we set ${\bf{N}}=50$. Figures (a), (b) and (c) show spins equal to 0.9, 0.99 and 0.999 respectively).
}\label{fig:FluxEccentrcity}
\end{figure*}

\begin{table*}[htbp]
  \centering
 \caption{ Computational cost {(floating-point numbers ${\bf N}$)} of GW energy fluxes for different Orbits.}  \label{tab:3orbits}
      \scalebox{0.92}{
          \begin{threeparttable}
    \begin{tabular}{c|c|cc|c|c|cc|c|c|cc|c|c|cc}
    \toprule
    \multicolumn{4}{c}{Circular Orbit$^\clubsuit$} & \multicolumn{4}{|c|}{Spherical Orbit$^\diamondsuit$} & \multicolumn{4}{|c|}{Eccentric Orbit$^\heartsuit$} & \multicolumn{4}{c}{Generic Orbit$^\spadesuit$} \\
    \midrule
    $a$    & $\Delta\dot{E}$ & HeunC & MST  & ${x_I}$   & $\Delta\dot{E}$ & HeunC & MST  &  $e$   & $\Delta\dot{E}$ & HeunC & MST  &  $e$   & $\Delta\dot{E}$ & HeunC & MST \\
    \midrule
    \multirow{2}[2]{*}{0.1} & $10^{-20}$   & 34   & 38   & \multirow{2}[2]{*}{0.1} & $10^{-20}$   & 34   & 39   & \multirow{2}[2]{*}{0.1} & $10^{-20}$   & 32   & 44   & \multirow{2}[2]{*}{0.1} & $10^{-20}$   & 32   & 43 \\
         & $10^{-30}$   & 44   & 56   &      & $10^{-30}$   & 44   & 61   &      & $10^{-30}$   & 42   & 64   &      & $10^{-30}$   & 42   & 63 \\
    \midrule
    \multirow{2}[2]{*}{0.3} & $10^{-20}$   & 36   & 38   & \multirow{2}[2]{*}{0.3} & $10^{-20}$   & 33   & 39   & \multirow{2}[2]{*}{0.3} & $10^{-20}$   & 35   & 44   & \multirow{2}[2]{*}{0.3} & $10^{-20}$   & 31   & 43 \\
         & $10^{-30}$   & 45   & 57   &      & $10^{-30}$   & 43   & 59   &      & $10^{-30}$   & 42   & 64   &      & $10^{-30}$   & 41   & 62 \\
    \midrule
    \multirow{2}[2]{*}{0.5} & $10^{-20}$   & 38   & 38   & \multirow{2}[2]{*}{0.5} & $10^{-20}$   & 33   & 39   & \multirow{2}[2]{*}{0.5} & $10^{-20}$   & 36   & 44   & \multirow{2}[2]{*}{0.5} & $10^{-20}$   & 32   & 43 \\
         & $10^{-30}$   & 48   & 57   &      & $10^{-30}$   & 44   & 59   &      & $10^{-30}$   & 42   & 65   &      & $10^{-30}$   & 42   & 63 \\
    \midrule
    \multirow{2}[2]{*}{0.7} & $10^{-20}$   & 43   & 43   & \multirow{2}[2]{*}{0.7} & $10^{-20}$   & 34   & 39   & \multirow{2}[2]{*}{0.7} & $10^{-20}$   & 35   & 42   & \multirow{2}[2]{*}{0.7} & $10^{-20}$   & 37   & 44 \\
         & $10^{-30}$   & 53   & 57   &      & $10^{-30}$   & 44   & 60   &      & $10^{-30}$   & 44   & 62   &      & $10^{-30}$   & 46   & 64 \\
    \midrule
    \multirow{2}[2]{*}{0.9} & $10^{-20}$   & 53   & 54   & \multirow{2}[2]{*}{0.9} & $10^{-20}$   & 34   & 40   & \multirow{2}[2]{*}{0.9} & $10^{-20}$   & 44   & 44   & \multirow{2}[2]{*}{0.9} & $10^{-20}$   & 39   & 46 \\
         & $10^{-30}$   & 64   & 78   &      & $10^{-30}$   & 44   & 61   &      & $10^{-30}$   & 49   & 63   &      & $10^{-30}$   & 47   & 68 \\
    \bottomrule
    \end{tabular}%
    \begin{tablenotes}
    \item[$\clubsuit$] The parameter of the circular orbit at the innermost stable circular orbit (ISCO) $p_{\rm ISCO}=6M$ is ${\ell _{{\rm{max}}}} = 6$.
    \item[$\diamondsuit$] The parameters of the spherical orbit at $p=10M$ are $a=0.9,e = 0,{\ell _{{\rm{max}}}} = 6, 1 \le n \le 5$.
    \item[$\heartsuit$] The parameters of the eccentric orbit at $p=10M$ are $a=0.9, {\theta _{{\rm{inc}}}} = 0, {\ell _{{\rm{max}}}} = 5 , 1 \le  k \le 3$.
    \item[$\spadesuit$] The parameters of the generic orbit at $p=10M$ are $a=0.9,{x_I} = 0.1,{\ell _{\max }} = 4,1 \le n \le 3,1 \le k \le 3$.
    \end{tablenotes}

    \end{threeparttable}

    }
\end{table*}

\begin{figure*}[htbp]
	\centering
\includegraphics[width=6in]{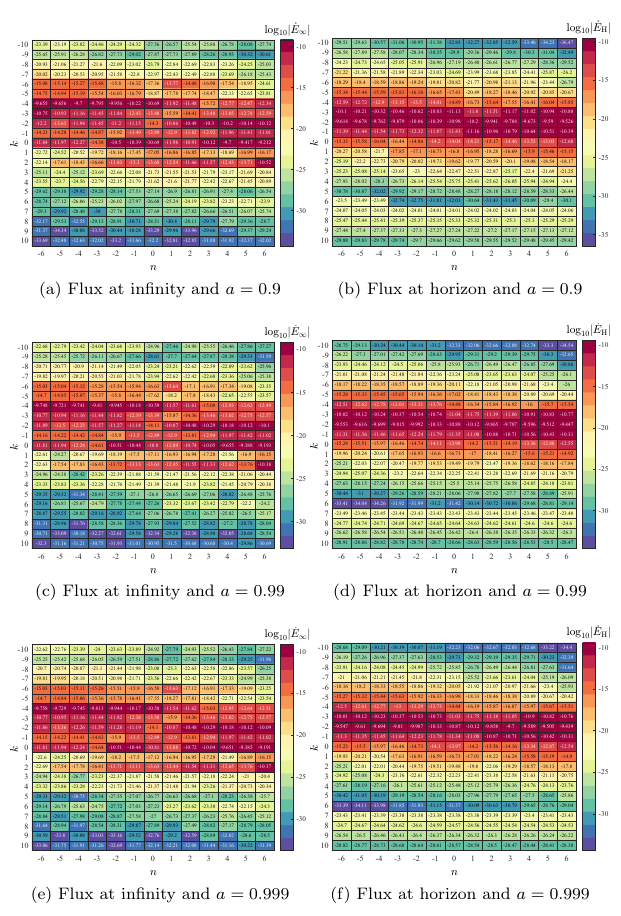}
\caption{Modal radiative fluxes ($\ell= m = 2$) through the horizon and at infinity for the generic orbit ($e = 0.9$ and $x_I=0.1$) at $p = 10M$ with large spins   $a =\{0.9,0.99,0.999\}$ and ${\bf{N}}=50$.}\label{fig:GOrbit-a-e}
\end{figure*}

To generate the data grids for energy flux, we need to calculate the energy flux corresponding to four orbital parameters $(a, p, e, x_I)$.
The resulting data grid, which stores these energy flux values, serves as an important database for constructing inspiral waveforms with any mass ratio.
It is estimated that 50\% of all observable EMRIs will have eccentricities $e > 0.2$ as they are close to the final stable orbit \cite{Barack:2003fp,Gair:2004iv,Drasco:2005kz}. Most EMRI events begin with large initial eccentricity, and calculating energy fluxes accurately and efficiently with large eccentricities is challenging for frequency-domain methods that use mode-sum regularization \cite{Wardell:2015kea}. In this section, we focus on the accuracy of radiative fluxes in scenarios involving high eccentricity, high spin, and strong-field regions.
We provide the relative errors of radiative fluxes corresponding to the four orbits using the three methods in \cref{tab:3orbits}.
{To demonstrate the high precision of our method, especially under conditions of high floating-point accuracy (${\bf N}=100$), we present error changes in radiative fluxes for both circular orbit and spherical orbits in \Cref{fig:FluxCircular,fig:FluxSpherical}.
}
From these numerical comparisons, it is evident that our method achieves better accuracy than the MST-NI method when using high floating-point precision. For radiative fluxes of eccentric orbits, their values show a special behavior as the index $n$ increases. They first rise to a peak value through oscillations and then attenuate, also oscillating \cite{Drasco:2005kz,Fujita:2009us}.
This behavior is illustrated in ~\cref{fig:FluxEccentrcity}.
Moreover, the results of the radiative fluxes of the generic orbital for large spins are shown in \cref{fig:GOrbit-a-e}.

Overall, our HeunC method demonstrates excellent applicability and high precision, making it suitable for calculating extreme cases involving general orbits with large eccentricities, higher spins, higher harmonic modes, and strong-field regions.
When calculating the radiative fluxes of single higher harmonic modes, it is important to ensure higher floating-point accuracy. The total energy flux, which is obtained by summing thousands of $(\ell, m, k, n)$ modes, plays a crucial role in constructing waveform templates. Our method achieves superior accuracy with lower floating-point numbers, which is advantageous for future efficiency enhancements.

\section{Conclusion}\label{sec:Conclusion}
In a previous study \cite{chen2023exact}, we used the confluent Heun function to calculate radiative fluxes resulting from the motion of test particles around a Schwarzschild BH in circular orbits.
This paper extends that work by calculating radiative fluxes generated by a particle in generic orbits around a near-extreme Kerr BH.
This extension is particularly relevant for modeling the gravitational radiation of binary star systems with varying mass ratios.
We begin by discussing the geodesic dynamics of a point particle on generic orbits in the Kerr spacetime.
Using the confluent Heun function, we then provide an exact solution to the Teukolsky equations for the Kerr spacetime. This solution allows us to determine gravitational wave amplitudes and radiative fluxes.
Finally, we compare our results with those obtained via numerical methods and find that our approach is highly satisfactory. The findings are summarized in detail as follows:

1. Our numerical simulations demonstrate that the HeunC method offers excellent applicability and high precision, even in extreme cases of general orbits with large eccentricities, higher spins, higher harmonic modes, and strong-field regions.
Most frequency domain methods \cite{Barton:2008eb,Wardell:2015kea,vandeMeent:2019cam} struggle to accurately compute GW radiative fluxes in strong-field regions with highly eccentric generic orbits (with eccentricities $e \gtrsim 0.5$).
Despite being a frequency domain approach, our method overcomes these computational challenges.

2. The MST method has proven successful in EMRI {waveform} calculations, exhibiting incomparable computational accuracy and efficiency \cite{Hughes:2021,Isoyama:2021jjd,skoup2023asymptotic}.
For generic orbits, calculating higher harmonic modes requires high floating-point accuracy owing to the involvement of four indexes $(\ell,m,k,n)$.
Our method, benefiting from the faster convergence speed of the HeunC function, surpasses both the MST-NI method and high-order post-Newtonian expansions in accuracy\footnote{In our previous study \cite{chen2023exact}, we have demonstrated in detail that the HeunC method exhibits a significantly faster convergence rate concerning floating-point numbers, compared to both the MST method, high-order post-Newtonian expansion, and numerical integration method.
}.
Ensuring higher floating-point accuracy is crucial when calculating radiative fluxes for single higher harmonic modes. The total energy flux, derived from summing thousands of $(\ell, m, k, n)$ modes, plays a crucial role in constructing waveform templates. Our method achieves superior accuracy using lower floating-point numbers, enhancing computational efficiency. {This instills confidence in us that the HeunC method and the self-consistent effective-one-body (EOB) theories \cite{Long:2024axi,Jing:2023vzq,Jing:2022vks,deng2024energy,Jing_2022} will be used for efficiently calculating GW waveforms in the future.}
\section*{Acknowledgement}

This work was supported by the Grant of NSFC Nos. 12035005 and 12122504,
and National Key Research and Development Program of China No. 2020YFC2201400.

\appendix


\section{Formulas for $E$, $L_z$, and $Q$}
\label{app:geodesicconstants}
We provide formulas for the geodesic constants of the motion $(E,L_z,Q)$ as functions of preferred parameters $(p,e,x_I)$.
The three geodesic constants of generic orbits can be written as
\begin{align}
 E &= \sqrt{\frac{\kappa\rho + 2\varpi\sigma - 2{\rm sgn}(x_I)\sqrt{\sigma(\sigma\varpi^2 + \rho\varpi\kappa - \eta\kappa^2)}}{\rho^2 + 4\eta\sigma}}\;,\label{eq:Eformula}\\
L_z &= -\frac{g(r_{\rm a})E - \sqrt{g(r_{\rm a})^2 + h(r_{\rm a})f(r_{\rm a})E^2 - h(r_{\rm a})d(r_{\rm a})}}{h(r_{\rm a})}\;,\label{eq:Lzformula}\\
Q &= (1 - x_I^2)\left[a^2(1 - E^2) + \left(\frac{L_z}{x_I}\right)^2 \right]\;, \label{eq:Qformula}
\end{align}
where
\begin{align}
  \kappa &= d(r_{\rm a})h(r_{\rm p}) - d(r_{\rm p})h(r_{\rm a})\;,
\\
\varpi &= d(r_{\rm a})g(r_{\rm p}) - d(r_{\rm p})g(r_{\rm a})\;,
\\
\rho &= f(r_{\rm a})h(r_{\rm p}) - f(r_{\rm p})h(r_{\rm a})\;,
\\
\eta &= f(r_{\rm a})g(r_{\rm p}) - f(r_{\rm p})g(r_{\rm a})\;,
\\
\sigma &= g(r_{\rm a})h(r_{\rm p}) - g(r_{\rm p})h(r_{\rm a})\;,
\end{align}
and
\begin{align}
 d(r) =& \Delta(r)[r^2 + a^2(1 - x_I^2)]\;,
\\
f(r) =& r^4 + a^2[r(r + 2M) + (1 - x_I^2)\Delta(r)]\;,
\\
g(r) =& 2aMr\;,
\\
h(r) =& r(r - 2M) + \frac{(1 - x_I^2)\Delta(r)}{x_I^2}\;,
\end{align}
where $r_{\rm a} = p/(1 - e)$ and $r_{\rm p} = p/(1 + e)$ are the coordinate radius of the orbit's apoapsis and periapsis, respectively.

More details of these formulas can be seen in Refs. \cite{Schmidt:2002qk,vandeMeent:2019cam}. In addition, we utilized some of Hughes's adjustments \cite{Hughes:2021} for the formula: $\varpi$ replaces $\epsilon$ in Refs. \cite{Schmidt:2002qk,vandeMeent:2019cam} to avoid the symbol conflict with $\epsilon = \sqrt{M^2 - a^2}/2Mr_+$ and the very similar $\varepsilon \equiv \mu/M$.

\section{Asymptotic Formula of HeunC Function at Infinity}\label{app:AsymptoticFormula}
In our previous study \cite{chen2023exact}, we derive the asymptotic analytic expression of the confluent Heun function at infinity. Bonelli et al. solved the connection problem of Heun class functions \cite{Bonelli:2022ten}. By using the connection formulae of $x\in[0,1]$ and $x\in[1,\infty]$ of the confluent Heun function derived by them, the connection formulae of $x\in[0,\infty]$ can be applied to calculate the HeunC value as $x$ tends to infinity. Our analytic expression can also be used to calculate the HeunC value as $x\rightarrow\infty$, and it also contains the amplitude of some physical information.
We provide the asymptotic expression of confluent Heun function $\mathbb{H} ={\rm{HeunC}}(\alpha ,\beta ,\gamma ,\delta ,\eta ;x)$ at infinity as follows
\begin{equation}
  \mathop {\lim }\limits_{|x| \to \infty }\mathbb{H}(x)= D_ \odot^\beta \;{x^{ - \frac{{\beta  + \gamma  + 2}}{2} - \frac{\delta }{\alpha }}} + D_\otimes ^\beta {{\rm{e}}^{ - \alpha x}}{x^{ - \frac{{\beta  + \gamma  + 2}}{2} + \frac{\delta }{\alpha }}} .
\end{equation}

Then, the constants $D_ \odot ^\beta$ and $D_ \otimes ^\beta$ are given by
\begin{align}
{}&\begin{array}{l}\label{eq:Dinc}
D_ \odot ^\beta  = \Xi _{n,\nu }^\beta D_{ \odot ,n,\nu }^\beta + {{\rm{e}}^{ - i\pi \left( {\nu  + \frac{1}{2}} \right)}}\frac{{\sin \pi \left( {\nu  + \frac{\delta }{\alpha }} \right)}}{{\sin \pi \left( {\nu  - \frac{\delta }{\alpha }} \right)}}\Xi _{ - n, - \nu  - 1}^\beta D_{ \odot , - n, - \nu  - 1}^\beta ,
\end{array}\\
{}&{D_ \otimes ^\beta  = \Xi _{n,\nu }^\beta D_{ \otimes ,n,\nu }^\beta  + {{\rm{e}}^{i\pi \left( {\nu  + \frac{1}{2}} \right)}}\Xi _{ - n, - \nu  - 1}^\beta D_{ \otimes , - n, - \nu  - 1}^\beta ,}\label{eq:Dout}
\end{align}
with
\begin{align}
D_{ \odot ,n,\nu }^\beta  &= {\left( { - 1} \right)^{\frac{{\gamma  + \beta  + 2}}{2} + \frac{\delta }{\alpha }}}{\big( {\frac{\alpha }{2}} \big)^\tau } {\big( -{\frac{{i\alpha }}{2}} \big)^{ - \frac{{\gamma  + \beta  + 2}}{2} - \frac{\delta }{\alpha }}}{{\rm{e}}^{ - \frac{{i\pi \tau +\alpha}}{2}}}
\nonumber \\
&\times
{2^{ - 1 - \frac{\delta }{\alpha }}}{{\rm{e}}^{\frac{{i\pi }}{2}\big( {\nu  + 1 + \frac{\delta }{\alpha }} \big)}} \Xi _{n,\nu }^\beta\frac{{\Gamma \big( {\nu  + 1 + \frac{\delta }{\alpha }} \big)}}{{\Gamma \big( {\nu  + 1 - \frac{\delta }{\alpha }} \big)}},
\\
 D_{  \otimes,n,\nu }^\beta  &= {\big( { - 1} \big)^{\frac{{\gamma  + \beta  + 2}}{2} - \frac{\delta }{\alpha }}}{\left( {\frac{\alpha }{2}} \right)^\tau } {\left(- {\frac{{i\alpha }}{2}} \right)^{ - \frac{{\gamma  + \beta  + 2}}{2} + \frac{\delta }{\alpha }}} \nonumber \\
&\times {{\rm{e}}^{ - \frac{{i\pi \tau - \alpha}}{2}}}\Xi _{n,\nu }^\beta  \frac{ {2^{ - 1 + \frac{\delta }{\alpha }}}{{\rm{e}}^{ - \frac{{i\pi }}{2}\big( {\nu  + 1 - \frac{\delta }{\alpha }} \big)}}} {\sum\limits_{n =  - \infty }^{ + \infty } {f_n^\nu } }   \nonumber \\
&\times\Big( {\sum\limits_{n =  - \infty }^{ + \infty } {{{( - 1)}^n}} \frac{{{{\big( {\nu  + 1 - \frac{\delta }{\alpha }} \big)}_n}}}{{{{\big( {\nu  + 1 + \frac{\delta }{\alpha }} \big)}_n}}}f_n^\nu } \Big).
\end{align}
and
\begin{align}\label{eq:Xinv}
&\begin{array}{*{20}{l}}
{\Xi _{n,\nu }^\beta  = \frac{{{2^{ - \nu }}{{\left( {\frac{\alpha }{2}} \right)}^{ - \hat \tau }}{{\rm{e}}^{\frac{{i\pi \hat \tau  + \alpha }}{2}}}\Gamma \left( {\beta  + 1} \right)\Gamma \left( {2\nu  + 2} \right)}}{{\Gamma \left( {\nu  + 1 + \frac{\delta }{\alpha }} \right)\Gamma \left( {\nu  + 1 - \frac{{\beta  + \gamma }}{2}} \right)\Gamma \left( {\nu  + 1 + \frac{{\gamma  - \beta }}{2}} \right)}}}\\
{ \times {{\Big( {\sum\limits_{n =  - \infty }^0 {\frac{{{{( - 1)}^n}{{\left( {\nu  + 1 - \frac{\delta }{\alpha }} \right)}_n}}}{{( - n)!{{\left( {2\nu  + 2} \right)}_n}{{\left( {\nu  + 1 + \frac{\delta }{\alpha }} \right)}_n}}}f_n^\nu } } \Big)}^{ - 1}}}
\end{array}\\
&\begin{array}{l}
{ \times \Big( {\sum\limits_{n = 0}^\infty  {{{\left( { - 1} \right)}^n}\frac{{\Gamma \left( {n + 2\nu  + 1} \right)\Gamma \left( {n + \nu  + 1 + \frac{{\gamma  - \beta }}{2}} \right)\Gamma \left( {n + \nu  + 1 - \frac{{\beta  + \gamma }}{2}} \right)}}{{( n!)\Gamma \left( {n + \nu  + 1 + \frac{{\beta  - \gamma }}{2}} \right)\Gamma \left( {n + \nu  + 1 + \frac{{\beta  + \gamma }}{2}} \right)}}f_n^\nu } } \Big)}.
\end{array}\nonumber
\end{align}
where $\tau  = \frac{1}{4}\left( {3\beta  + \gamma  + \frac{{2\delta }}{\alpha }} \right)$, and $f_n^\nu$ satisfy the following three-term recurrence relation,
\begin{equation}\label{recurrence-fnv}
 {{\hat \alpha }_n}f_{n + 1}^\nu  + {{\hat \beta }_n}f_n^\nu  + {{\hat \gamma }_n}f_{n - 1}^\nu  = 0,
\end{equation}
where
\begin{subequations}
\begin{align}
{{\hat \alpha }_n} &=   \frac{{\left( {2n + 2\nu  + 2 - \beta  + \gamma } \right)}}{{8\left( {n + \nu  + 1} \right)\left( {2n + 2\nu  + 3} \right)}}\nonumber \\
&\times \left( {\alpha n + \alpha \nu  + \alpha  -\delta } \right)\left( {2n + 2\nu  + 2 - \gamma  - \beta } \right),\\
{{\hat \beta }_n} &= \eta  + \frac{\delta }{2} - \frac{{{\beta ^2}}}{4} - \frac{{{\gamma ^2}}}{4} + \left( {n + \nu } \right)\left( {n + \nu  + 1} \right) \nonumber \\
&+ \frac{{\delta \left( {\gamma  + \beta } \right)\left( {\beta  - \gamma } \right)}}{{8\left( {n + \nu } \right)\left( {n + \nu  + 1} \right)}},\\
{{\hat \gamma }_n} &= -\frac{{\left( {2n + 2\nu  + \beta  - \gamma } \right)}}{{8\left( {n + \nu } \right)\left( 2n + 2\nu  - 1 \right)}} \nonumber \\
&\times \left( {\alpha n + \alpha \nu  + \delta } \right)\left( {2n + 2\nu  + \gamma  + \beta } \right).
\end{align}
\end{subequations}

The phase parameter $\nu$, also known as the renormalized angular momentum, may be obtained by solving a characteristic equation expressed as the sum of two infinite continued fractions.
\begin{equation}\label{eq:nu}
  {{\hat \beta }_0} = \frac{{{{\hat \alpha }_{ - 1}}{{\hat \gamma }_0}}}{{{{\hat \beta }_{ - 1}} - }}\frac{{{{\hat \alpha }_{ - 2}}{{\hat \gamma }_{ - 1}}}}{{{{\hat \beta }_{ - 2}} - }}\frac{{{{\hat \alpha }_{ - 3}}{{\hat \gamma }_{ - 2}}}}{{{{\hat \beta }_{ - 3}} - }} +  \cdots
 + \frac{{{{\hat \alpha }_0}{{\hat \gamma }_1}}}{{{{\hat \beta }_1} - }}\frac{{{{\hat \alpha }_1}{{\hat \gamma }_2}}}{{{{\hat \beta }_2} - }}\frac{{{{\hat \alpha }_2}{{\hat \gamma }_3}}}{{{{\hat \beta }_3} - }} \cdots .
\end{equation}



\section{Parameters of HeunC Functions}\label{app:HeunCParameters}
Currently, three primary mathematical software packages offer numerical calculation capabilities for the confluent Heun function, as well as for its derivation and integration. {Maple} and Mathematica provide official built-in functions for these calculations, while the calculation code of MATLAB is provided by Motygin \cite{Motygin_2015,Motygin2018}.

For Maple software, the ODE of the confluent Heun function $\mathbb{H}_{\rm{0}}^{\beta } (x) = {\rm{HeunC}}(\alpha, \beta,\gamma,\delta,\eta;x) $ can be written as
\begin{align}\label{eq:HeunC-Maple}
&\mathbb{H} '' - \frac{{\left( { - {x^2}\alpha  + \left( { - 2 - \beta  - \gamma  + \alpha } \right)x + 1 + \beta } \right)}}{{x\left( {x - 1} \right)}}\mathbb{H}'\nonumber\\
 &  - \big{[} \left( {\left( { - 2 - \beta  - \gamma } \right)\alpha  - 2\delta } \right)x
 + \left( {\beta  + 1} \right)\alpha
\nonumber\\
 &    + \left( { - \gamma  - 1} \right)\beta  - \gamma  - 2\eta \big{]}\frac{\mathbb{H} }{{2x\left( {x - 1} \right)}} = 0.
\end{align}
For MATLAB/Mathematica software, the ODE of the confluent Heun function can be written as
\begin{equation}\label{eq:HeunC-MMA}
 {\bf{H''}} + {\bf{H'}}\left( {\frac{{{\gamma _{\rm{M}}}}}{x} + \frac{{{\delta _{\rm{M}}}}}{{x - 1}} + {\epsilon _{\rm{M}}}} \right) + \frac{{({\alpha _{\rm{M}}}x - {q_{\rm{M}}})}}{{x(x - 1)}}{\bf{H}} = 0
\end{equation}
where ${\bf{H}}(x) = {\rm{HeunC}}({q_{\rm{M}}},{\alpha _{\rm{M}}},{\gamma _{\rm{M}}},{\delta _{\rm{M}}},{\epsilon _{\rm{M}}};x)$.

The parameter relation between \Cref{eq:HeunC-Maple,eq:HeunC-MMA} is as follows
\begin{subequations}
\begin{align}
  {q_{{\rm{M}}}} &= \frac{1}{2}\alpha \beta  - \frac{1}{2}\beta \gamma  + \frac{1}{2}\alpha  - \frac{1}{2}\beta  - \eta  - \frac{1}{2}\gamma ,\\
{\alpha _{{\rm{M}}}} &= \frac{1}{2}\alpha \beta  + \frac{1}{2}\alpha \gamma  + \alpha  + \delta ,\\
{\delta _{{\rm{M}}}} &= \gamma  + 1,\\
{\epsilon _{{\rm{M}}}} &= \alpha ,\\
{\gamma _{{\rm{M}}}} &= \beta  + 1.
\end{align}
\end{subequations}

\bibliographystyle{scpma-zycai} 
\bibliography{mybibfile}

\end{multicols}

\end{document}